\documentclass[%
aip,
rsi,
amsmath,amssymb,
reprint,%
]{revtex4-2}

\usepackage[utf8]{inputenc}
\usepackage[T1]{fontenc}
\usepackage[german, english]{babel}
\usepackage[version=4]{mhchem}
\usepackage{graphicx}% Include figure files
\usepackage{units}
\usepackage{textcomp}
\usepackage[Omega]{gensymb}
\usepackage{xspace}
\usepackage[usenames,dvipsnames]{xcolor}
\usepackage{soul}

\newcommand{\eg}{\mbox{e.g.\ }}

\newcommand{\msub}[1]{_{\text{#1}}}

\newcommand{\refF}[1]{Fig.\,\ref{#1}}
\newcommand{\refT}[1]{Tab.\,\ref{#1}}
\newcommand{\refE}[1]{Eq.\,\ref{#1}}

\newcommand{\diff}{\mathop{}\!\mathrm{d}}
\newcommand{\kb}{k\msub{B}}
\newcommand{\rs}{R\msub{S}}
\newcommand{\trs}{$R\msub{S}$\xspace} % text-version of R_S
\newcommand{\rres}{R\msub{res}}
\newcommand{\tres}{$R\msub{res}$\xspace} % text-version of Rres
\newcommand{\rbcs}{R\msub{BCS}}
\newcommand{\tbcs}{$R\msub{BCS}$\xspace} % text-version of Rres

\begin{document}

\title{Mitigation of parasitic losses in the quadrupole resonator enabling direct measurements of low residual resistances of SRF samples}

\author{S.~Keckert}
    \email[]{sebastian.keckert@helmholtz-berlin.de}
    \affiliation{Helmholtz-Zentrum Berlin, Hahn-Meitner-Platz 1, 14109 Berlin, Germany}
\author{W.~Ackermann}	\affiliation{Technische Universit{\"a}t Darmstadt, Schlo{\ss}gartenstra{\ss}e 8, 64289 Darmstadt, Germany}
\author{H.~De~Gersem}	\affiliation{Technische Universit{\"a}t Darmstadt, Schlo{\ss}gartenstra{\ss}e 8, 64289 Darmstadt, Germany}
\author{X.~Jiang}		\affiliation{Universit{\"a}t Siegen, Adolf-Reichwein-Stra{\ss}e 2a, 57076 Siegen, Germany}
\author{A.~{\"O}.~Sezgin} \affiliation{Universit{\"a}t Siegen, Adolf-Reichwein-Stra{\ss}e 2a, 57076 Siegen, Germany}
\author{M.~Vogel}		\affiliation{Universit{\"a}t Siegen, Adolf-Reichwein-Stra{\ss}e 2a, 57076 Siegen, Germany}
\author{M.~Wenskat}		\affiliation{Universit{\"a}t Hamburg, Mittelweg 177, 20148 Hamburg, Germany}
\author{R.~Kleindienst}	\affiliation{Helmholtz-Zentrum Berlin, Hahn-Meitner-Platz 1, 14109 Berlin, Germany}
\author{J.~Knobloch}
    \affiliation{Helmholtz-Zentrum Berlin, Hahn-Meitner-Platz 1, 14109 Berlin, Germany}
    \affiliation{Universit{\"a}t Siegen, Adolf-Reichwein-Stra{\ss}e 2a, 57076 Siegen, Germany}
\author{O.~Kugeler}		\affiliation{Helmholtz-Zentrum Berlin, Hahn-Meitner-Platz 1, 14109 Berlin, Germany}
\author{D.~Tikhonov}	\affiliation{Helmholtz-Zentrum Berlin, Hahn-Meitner-Platz 1, 14109 Berlin, Germany}

\date{\today}

\begin{abstract}
The quadrupole resonator (QPR) is a dedicated sample-test cavity for the RF characterization of superconducting samples in a wide temperature, RF field and frequency range.
Its main purpose are high resolution measurements of the surface resistance with direct access to the residual resistance thanks to the low frequency of the first operating quadrupole mode.
Besides the well-known high resolution of the QPR, a bias of measurement data towards higher values has been observed, especially at higher harmonic quadrupole modes.
Numerical studies show that this can be explained by parasitic RF losses on the adapter flange used to mount samples into the QPR.
Coating several micrometer of niobium on those surfaces of the stainless steel flange that are exposed to the RF fields significantly reduced this bias, enabling a direct measurement of a residual resistance smaller than \unit[5]{n\ohm} at \unit[2]{K} and \unit[413]{MHz}.
A constant correction based on simulations was not feasible due to deviations from one measurement to another.
However, this issue is resolved given these new results.
\end{abstract}

% insert suggested PACS numbers in braces on next line

%\maketitle must follow title, authors, abstract and \pacs
\maketitle

\section{Introduction} \label{sec:intro}
Superconducting radio frequency (SRF) cavities are a key component for many state-of-the-art particle accelerators.
One important figure of merit of such a cavity is its quality factor, determined by its geometry and by the material's surface resistance.
The surface resistance as a function of RF frequency $(f)$ and temperature $(T)$ is composed of several contributions, commonly approximated by
\begin{equation} \label{eq:rs}
	\rs = \rbcs + \rres = \frac{af^2}{T}\exp\left(-\frac{\Delta}{\kb T}\right) + \rres
\end{equation}
with intrinsic BCS resistance $(\rbcs)$ and a residual resistance $(\rres)$.
Both equations are valid at low fields only and do not contain effects like field-dependent losses.
\tbcs depends on a material parameter $(a)$ and the superconducting energy gap $(\Delta)$. The contributions to \tres are less well understood and are still under investigation.
Hence, for R\&D on materials, coatings or surface treatments aiming at application in SRF cavities, precision measurements of the surface resistance are required.

\begin{figure}[htb]
	\centering
	\includegraphics[width=\columnwidth]{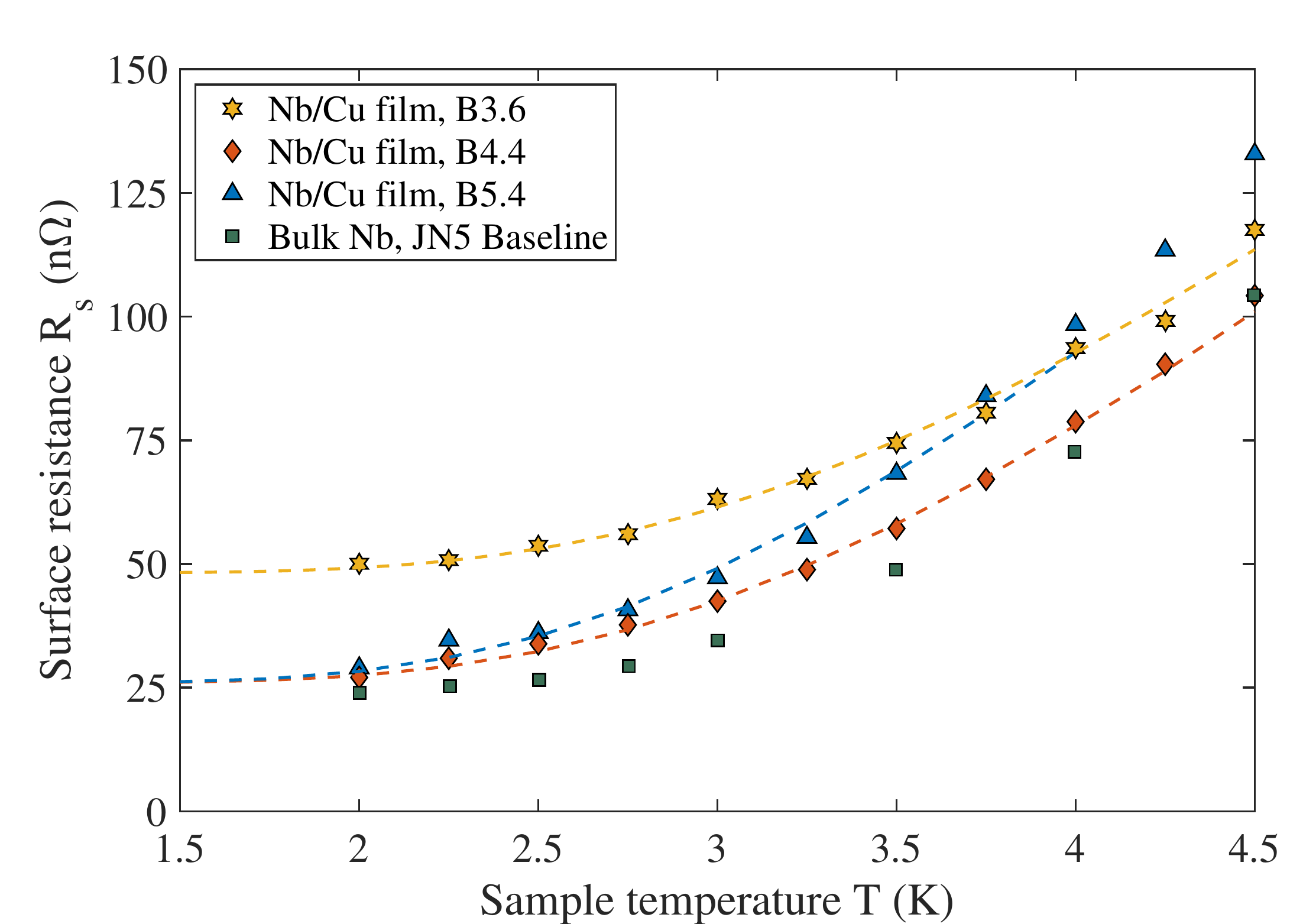}
	\caption{Representative measurement results of different niobium on copper films as well as for a bulk Nb sample at \unit[415]{MHz}. The BCS resistance is fitted using \refE{eq:rs}.}
	\label{fig:example-data}
\end{figure}

The Quadrupole Resonator (QPR) is a dedicated sample-test cavity, providing high resolution measurements in a wide parameter space of temperature and RF field at three different frequencies \cite{qpr_cern_1998_epac_construction, qpr_cern_2003_rsi, qpr_cern_2012_rsi, qpr_hzb_2021_rsi}.
With a first operating quadrupole mode at about \unit[415]{MHz}, \tbcs at \unit[2]{K} is typically smaller than \unit[2]{n\ohm}, enabling direct measurements of \tres.
Operational experience of the QPR indicates a bias of measurement data towards a systematically overestimated \trs, limiting the measurement accuracy at low \trs and hence especially impacting \tres \cite{diss_raphael, diss_sebastian, qpr_hzb_2018_ipac}.
As an example, \refF{fig:example-data} shows a series of measurements of \trs vs.\ temperature for different Nb on copper films as well as for a bulk Nb sample (JN5).
The results show that at the first quadrupole mode $(f\approx\unit[415]{MHz})$, \tres is larger than \unit[20]{n\ohm}, even for the bulk niobium sample.

The QPR uses a calorimetric compensation technique to derive the surface resistance of the sample of interest.
First, the sample is heated to a temperature of interest without RF field and in thermal equilibrium the required heater power is recorded.
Then, the RF is switched on to the field amplitude of interest.
A control loop reduces the DC heater power to reach again and stabilize the temperature of interest.
In thermal equilibrium the RF dissipated power is given by the difference of heater power, allowing to calculate the surface resistance according to

\begin{equation} \label{eq:rs-calc}
	\rs = \frac{2 P\msub{diss}}{\int\msub{sample}||H||^2 \diff S} = 2c \frac{\Delta P\msub{DC}}{P_t Q_t}
\end{equation}
with transmitted power $P_t$ measured at the pickup antenna with coupling $Q_t$.
The calibration constant $c$ is known from simulations, giving the ratio of stored energy to the integral of RF field on the sample surface.
With this calorimetric compensation technique, any heating occurring in the thermal system of the sample assembly is interpreted as surface resistance of the sample.

In the further course of this paper we show that the observed behavior of biased residual resistance can be explained by parasitic losses on normal conducting parts of the sample chamber assembly.
We provide measurement data of a niobium film sample tested first using a standard stainless steel adapter flange and exhibiting a residual resistance of about \unit[29]{n\ohm} at the first quadrupole mode.
The situation is exacerbated at \unit[1.3]{GHz} due to a reduced damping in the coaxial gap and a measurement of the residual resistance hitherto was impossible.

Applying a superconducting coating to the flange surfaces that are exposed to (small) RF fields yielded \tres of less than \unit[5]{n\ohm}.
For the first time, QPR measurements at a frequency near \unit[1.3]{GHz} gave $\rs<\unit[35]{n\ohm}$, enabling to investigate the frequency-dependent residual resistance.
The observed reduction of measured \trs is in good agreement with numerical simulations, proving a significant suppression of the bias and boosting the accuracy of measurement data to an unprecedented level.

\section{Numerical analysis of parasitic losses} \label{sec:simulations}
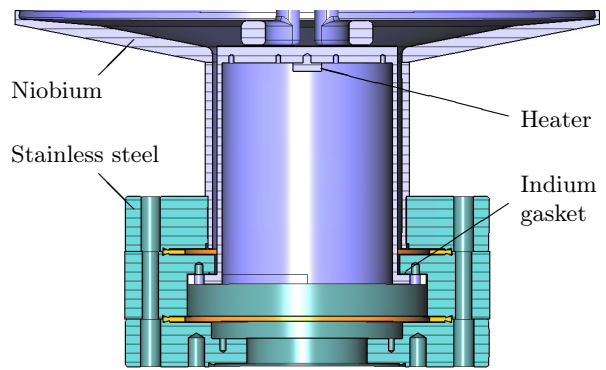
\begin{figure}[htb]
	\centering
	\setlength{\unitlength}{0.9\columnwidth}
	\includegraphics[width=0.9\columnwidth]{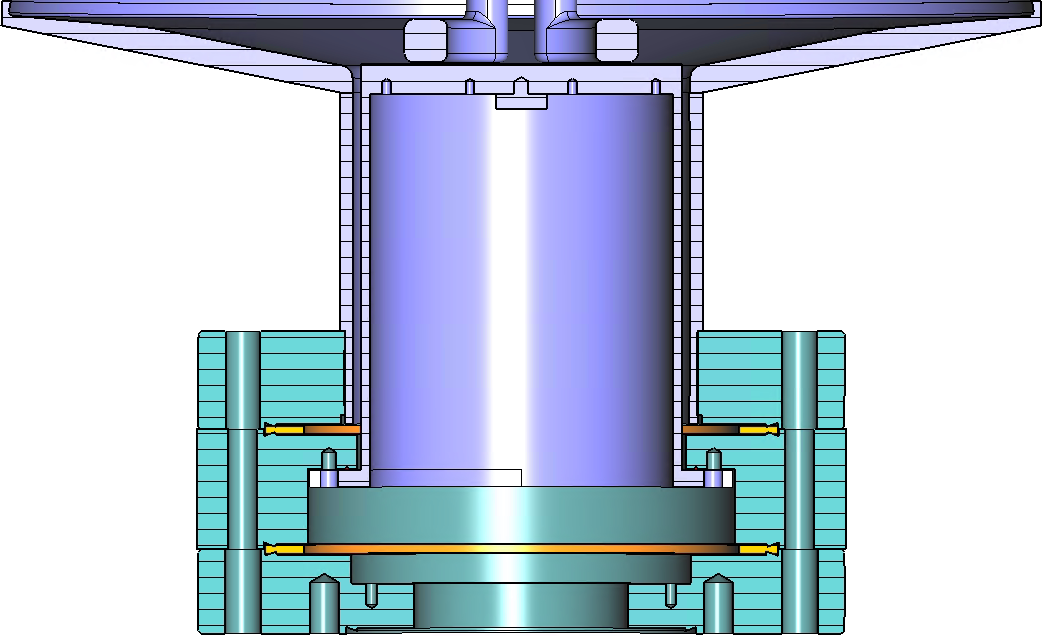}
	\begin{picture}(0,0)(1,0)
		\put(-0.02,0.46){Niobium}
		\put(-0.02,0.35){Stainless steel}
		\put( 0.13,0.50){\line(1, 1){0.06}}
		\put( 0.13,0.33){\line(1,-1){0.06}}
		\put( 0.85,0.41){Heater}
		\put( 0.83,0.43){\line(-4, 1){0.32}}
		\put( 0.85,0.30){Indium}
		\put( 0.85,0.25){gasket}
		\put( 0.83,0.28){\line(-3,-2){0.175}}
	\end{picture}
	\caption{Cross section of the sample chamber assembly mounted into the lower part of the QPR. The actual sample surface is given by the circular top area positioned at close distance to the QPR pole shoes. A heater and several temperature sensors (not depicted) are positioned underneath the surface of the sample.}
	\label{fig:simulation-geometry}
\end{figure}

In the following, the QPR design is described only briefly. For further details the reader is referred to \cite{qpr_hzb_2021_rsi}. The QPR sample chamber assembly consists of a top-hat-shaped superconducting part and a stainless steel adapter flange as shown in \refF{fig:simulation-geometry}.

The cylindrical sidewall of the inner part is manufactured of high RRR bulk niobium. For R\&D on thin films, the top disk acting as a substrate and carrying the sample surface is either build from bulk niobium or copper. The inner part and adapter flange are detachable. An indium wire provides the vacuum seal separating the inner insulation vacuum of the sample chamber from the resonator volume. The assembled sample chamber is inserted into the QPR from below, acting as an inner conductor of a coaxial line as illustrated in \refF{fig:simulation-geometry}.

Within the entire resonating structure, a particular mode has to be excited such that a high magnetic field strength will be available on the surface of the sample. For this purpose, a dedicated input coupler is mounted at the ceiling of the resonator where an external power source can be connected. Due to the structural design of the internal system with four rods and two horse-shoe interconnects a quadrupole mode resonating at approximately \unit[431.7]{MHz} serves as lowest-order mode. Corresponding higher-order modes with repetitive fields along the rods then oscillate at \unit[868.9]{MHz} and at \unit[1311.7]{MHz} with nearly identical fields in the vicinity of the sample surface. Especially the first and the last mentioned modes are of interest for many international cavity-design projects because of the proximity to their operational frequencies. Recognizable variations between the measured and the simulated frequencies can be attributed to an idealized numerical model. In this study, all numerical simulations have been performed with the help of the CST Studio Suite \cite{CSTStudioSuite}.

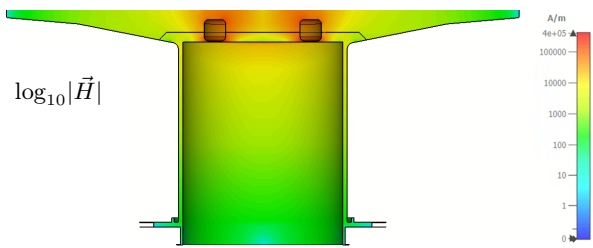
\begin{figure}[htb]
	\centering
	\setlength{\unitlength}{0.9\columnwidth}
	\includegraphics[width=0.9\columnwidth]{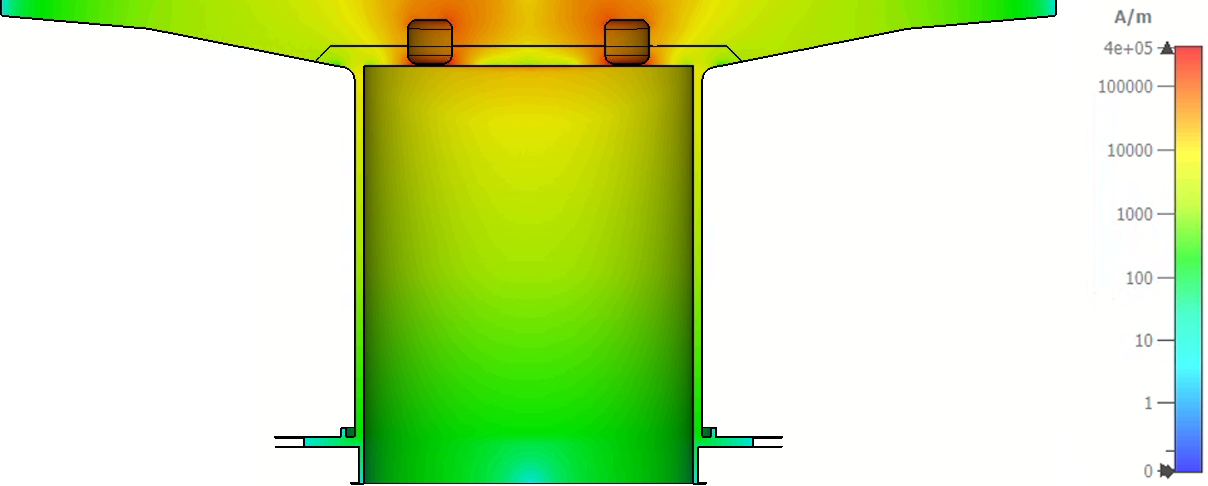}
	\begin{picture}(0,0)(1,0)
		\put(0,0.25){$\textrm{log}_{10}|\vec{H}|$}
	\end{picture}
	\caption{Simulated distribution of the magnetic field strength in a cross section of the lower part of the QPR.}
	\label{fig:simulation-field-mag}
\end{figure}

Unfortunately, the desired high magnetic field strength at the top of the sample surface comes together with high magnetic fields in the gap between the sample holder and the lower part of the cavity. This oscillating high magnetic field will subsequently induce related surface currents which will heat up the surface according to the deposited power. In order to attenuate this parasitic electromagnetic fields within the coaxial structure the relevant waveguide modes have to be excited below their cutoff frequencies such that they are exponentially decaying towards the end of the line. For the first relevant quadrupole mode in the cavity the simulated magnetic field strength is displayed in the vicinity of the sample surface as well as in the coaxial gap in \refF{fig:simulation-field-mag}. Here, a logarithmic scaling is used to demonstrate the deep penetration into the coaxial line.

\begin{figure}[htb]
	\centering
	\setlength{\unitlength}{0.7\columnwidth}
	\includegraphics[width=0.7\columnwidth]{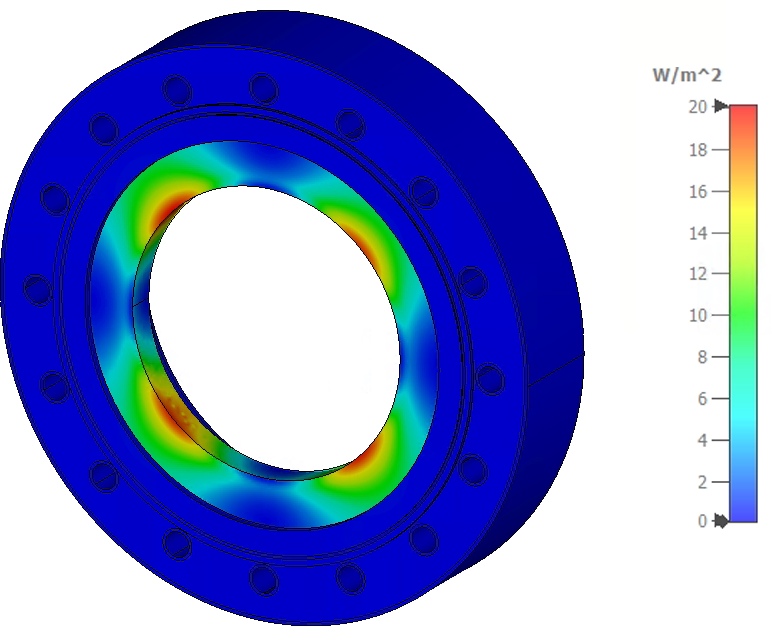}
	\begin{picture}(0,0)(1,0)
		\put(-0.21,0.74){Flange}
		\put(-0.04,0.72){\line(1,-1){0.1}}
	\end{picture}
	\caption{Simulated distribution of the surface power density on the flange associated to the sample.}
	\label{fig:simulation-surface-power}
\end{figure}
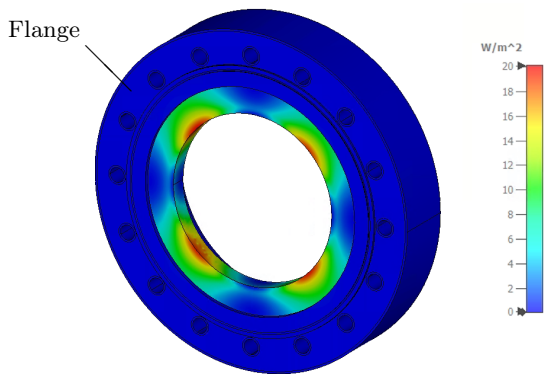

According to the simulated magnetic field strength an unavoidable surface current will be induced on the cylindrical walls of the coaxial line and on the stainless steel flanges as well as on the copper and indium seals. To estimate the contribution of the parasitic heating originating merely from the mounting of the probe into the resonator the electrical conductivity of niobium at cryogenic temperatures is assumed to be infinitely high compared to the finite values of the surrounding materials specified by $\sigma\msub{Copper}=\unit[3.3\!\times\!10^9]{S/m}$, $\sigma\msub{Indium}=\unit[3.6\!\times\!10^8]{S/m}$ and $\sigma\msub{Steel}=\unit[1.8\!\times\!10^6]{S/m}$ \cite{material_properties_fermilab}. The induced surface current density acting on a material with a finite electrical conductivity will result in a surface power density which finally will locally heat up the field-exposed components to possibly unacceptable temperatures. The simulated distribution of the surface power density on the sample adapter flange is visualized in \refF{fig:simulation-surface-power} where the scaling of the field is chosen such that the total stored energy in the system is given by \unit[1]{J}.

The unwanted quadrupolar-like power-density distribution on the surface of the flange is naturally originating from the operating quadrupole mode within the QPR. This parasitic power density is numerically determined with the help of eigenvalue calculations and has to be scaled accordingly within the subsequent thermal calculations where different magnetic field amplitudes have to be considered.

Due to the nonlinear thermal behavior of the underlying materials, the simulation of the steady-state temperature distribution has to be restarted again and again once the magnitude of the excitation has changed. A feedback loop back to the eigenvalue solver is not required because the electric conductivities of the applied materials are considered to be constant in the occurring temperature range.
Thermal conductivities of the involved materials are taken from \cite{niob_thermal_conductivity, nist_cryogenic_properties}.

\begin{figure}[htb]
	\centering
	\setlength{\unitlength}{0.9\columnwidth}
	\includegraphics[width=0.9\columnwidth]{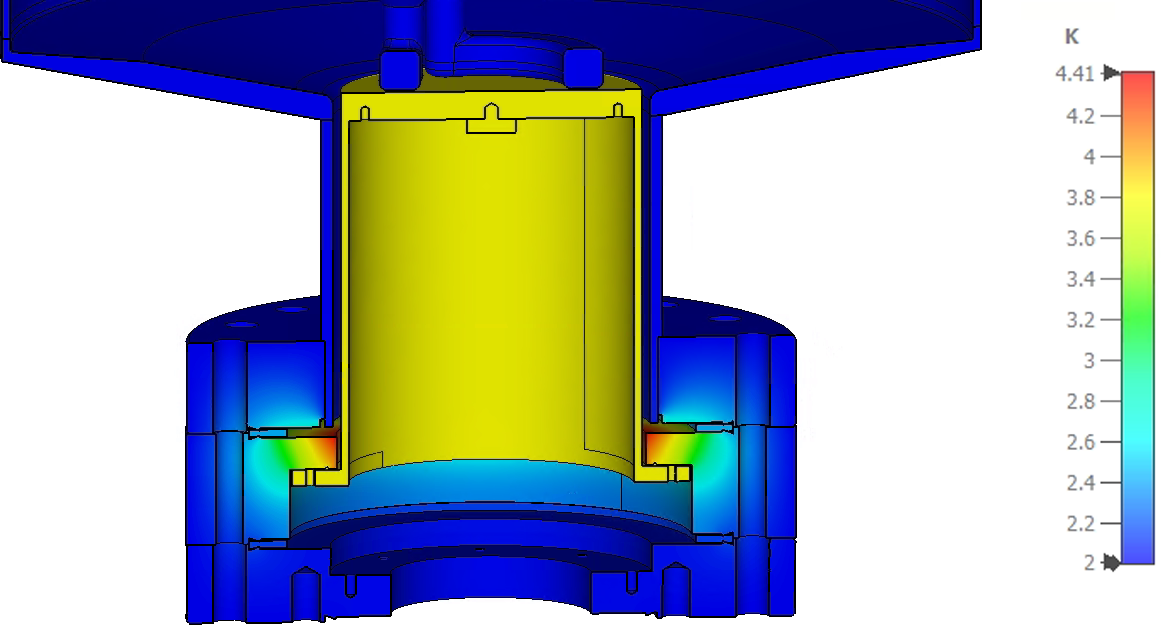}
	\begin{picture}(0,0)(1,0)
		\put(-0.02,0.22){Flange}
		\put(-0.02,0.33){Cavity}
		\put( 0.08,0.20){\line(1,-1){0.08}}
		\put( 0.08,0.37){\line(1, 1){0.08}}
	\end{picture}
	\caption{Simulated temperature distribution due to power dissipation on finite electrical conductive materials (stainless steel flanges as well as the copper and indium seals) in the lower part of the QPR.}
	\label{fig:simulation-temperature-mode}
\end{figure}
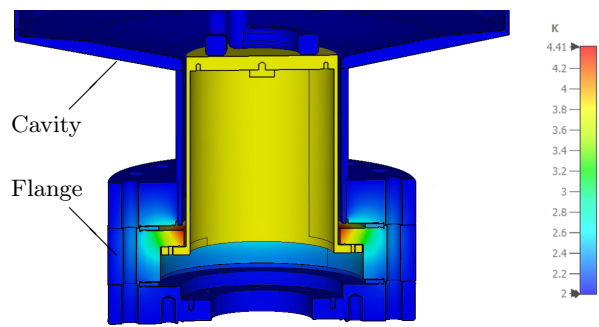

For a given field intensity, the calculated temperature distribution in the vicinity of the sample and the flanges is visualized in \refF{fig:simulation-temperature-mode}. Here, the cut plane is chosen such that the hot spots of the temperature due to the quadrupolar excitation become apparent. The highest temperatures naturally arise where the power density is maximal and decrease towards the background which is modeled by defining a fixed-temperature boundary condition of \unit[2]{K}. During operation, the entire system is immersed in a bath of superfluid helium which perfectly justifies the assumed numerical isothermal boundary condition.

On account of the fact that the thermal conductivity of niobium is much higher than the one of stainless steel the entire inner cylinder exhibits a nearly constant temperature. Since this sample temperature is determined not only by the applied RF field but also by the chosen geometries and materials, the temperature value itself is not a representative measure to compare different flange designs.

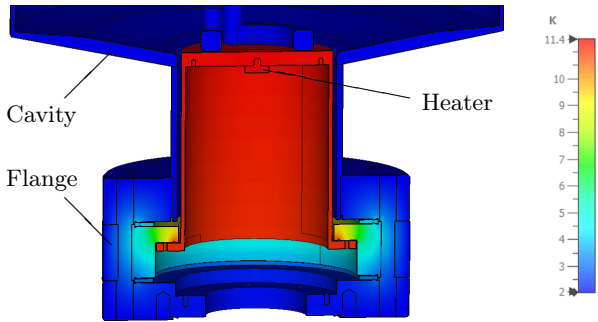
\begin{figure}[htb]
	\centering
	\setlength{\unitlength}{0.9\columnwidth}
	\includegraphics[width=0.9\columnwidth]{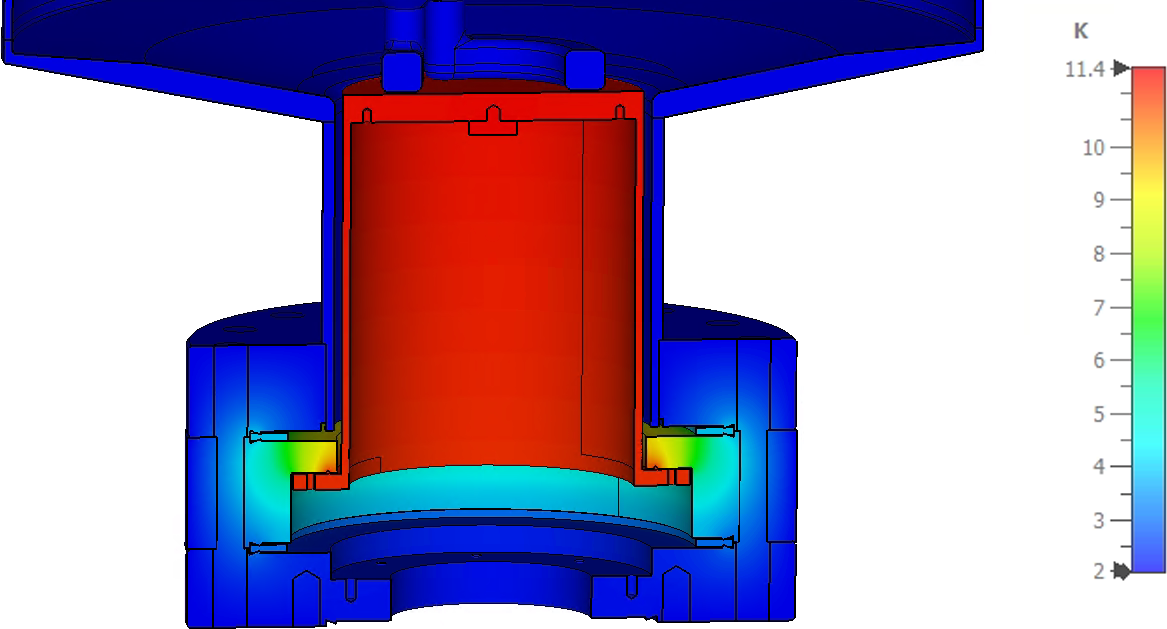}
	\begin{picture}(0,0)(1,0)
		\put(-0.02,0.23){Flange}
		\put(-0.02,0.34){Cavity}
		\put( 0.69,0.36){Heater}
		\put( 0.08,0.21){\line( 1,-1){0.08}}
		\put( 0.08,0.38){\line( 1, 1){0.08}}
		\put( 0.67,0.38){\line(-5, 1){0.25}}
	\end{picture}
	\caption{Simulated temperature distribution in the lower part of the QPR based on an intentional power insertion using a dedicated heater underneath the sample.}
	\label{fig:simulation-temperature-heater}
\end{figure}

For this reason, additional thermal calculations with various power excitations on a dedicated heater underneath the sample surface have been accomplished. The resulting sample temperatures as a function of the excitation power can then reliably be used to specify an effective heating power to each simulated sample temperature independent of the actual heat source. In \refF{fig:simulation-temperature-heater} the result of one of those calculations is visualized where the power is incorporated into the model by a dedicated volumetric heating device underneath the probe.

Following this idea, a series of such calculations does not only enable to assign an effective heating power to each calculated steady-state sample temperature, it also allows to quantify the influence of the artificial heating with the help of an additional surface resistance originating from parasitic heating effects. In contrast to the resulting steady-state sample temperature or the effective heating power this quantity directly allows to estimate the influence of the parasitic heating in terms of the surface resistance under investigation.

\begin{table*}
	\caption{Sample temperature, effective power and parasitic surface resistance for the first three quadrupole modes of the QPR resulting from the finite conductivity of the applied materials.}
	\label{tab:sim-data-flange}
	\begin{tabular}{|c|c|c|c|c|c|c|c|c|c|}
		\hline
		& \multicolumn{3}{c|}{$Q_1$} & \multicolumn{3}{c|}{$Q_2$} & \multicolumn{3}{c|}{$Q_3$}\\\hline
		$\;B(\unit[]{mT})\;$ & $\;T(\unit[]{K})\;$ & $\;P(\unit[]{mW})\;$ & $\;\rs(\unit[]{n\ohm})\;$ & $\;T(\unit[]{K})\;$ & $\;P(\unit[]{mW})\;$ & $\;\rs(\unit[]{n\ohm})\;$ & $\;T(\unit[]{K})\;$ & $\;P(\unit[]{mW})\;$ & $\;\rs(\unit[]{n\ohm})\;$\\\hline
		$  5$ & $2.008$ & $0.067$ & $12.6$ & $2.015$ & $0.129$ & $26.0$ & $2.032$ & $0.282$ & $63.1$\\\hline
		$ 10$ & $2.030$ & $0.266$ & $12.6$ & $2.058$ & $0.515$ & $26.0$ & $2.124$ & $1.127$ & $63.0$\\\hline
		$ 20$ & $2.117$ & $1.062$ & $12.6$ & $2.219$ & $2.050$ & $25.9$ & $2.446$ & $4.449$ & $62.1$\\\hline
		$ 50$ & $2.621$ & $6.527$ & $12.4$ & $3.059$ & $12.84$ & $25.9$ & $3.879$ & $30.08$ & $67.3$\\\hline
		$100$ & $3.803$ & $28.17$ & $13.4$ & $4.801$ & $58.56$ & $29.5$ & $6.513$ & $131.6$ & $73.6$\\\hline
	\end{tabular}
\end{table*}

\begin{table*}
	\caption{Sample temperature, effective power and parasitic surface resistance for the first three quadrupole modes of the QPR resulting from the finite conductivity of the applied materials. The sample adapter flange has been coated with niobium.}
	\label{tab:sim-data-flange-sputter}
	\begin{tabular}{|c|c|c|c|c|c|c|c|c|c|}
		\hline
		& \multicolumn{3}{c|}{$Q_1$} & \multicolumn{3}{c|}{$Q_2$} & \multicolumn{3}{c|}{$Q_3$}\\\hline
		$\;B(\unit[]{mT})\;$ & $\;T(\unit[]{K})\;$ & $\;P(\unit[]{mW})\;$ & $\;\rs(\unit[]{n\ohm})\;$ & $\;T(\unit[]{K})\;$ & $\;P(\unit[]{mW})\;$ & $\;\rs(\unit[]{n\ohm})\;$ & $\;T(\unit[]{K})\;$ & $\;P(\unit[]{mW})\;$ & $\;\rs(\unit[]{n\ohm})\;$\\\hline
		$  5$ & $2.000$ & $0.001$ & $0.26$ & $2.000$ & $0.003$ & $0.54$ & $2.001$ & $0.006$ & $1.36$\\\hline
		$ 10$ & $2.001$ & $0.005$ & $0.26$ & $2.001$ & $0.011$ & $0.54$ & $2.003$ & $0.024$ & $1.36$\\\hline
		$ 20$ & $2.002$ & $0.022$ & $0.26$ & $2.005$ & $0.043$ & $0.54$ & $2.011$ & $0.097$ & $1.36$\\\hline
		$ 50$ & $2.016$ & $0.136$ & $0.26$ & $2.030$ & $0.269$ & $0.54$ & $2.068$ & $0.606$ & $1.35$\\\hline
		$100$ & $2.061$ & $0.544$ & $0.26$ & $2.118$ & $1.071$ & $0.54$ & $2.255$ & $2.408$ & $1.35$\\\hline
	\end{tabular}
\end{table*}

In a parametric study, a number of thermal calculations have been performed where the magnitude of the RF field within the QPR has been varied. The results of the comprehensive simulation campaign are summarized in \refT{tab:sim-data-flange}. In this setup, the external heater is not used to heat up the probe in advance to a predefined temperature level. The required electromagnetic field distribution is obtained with the help of a numerical eigenmode calculation where the calculated fields are scaled such that the specified maximum magnetic flux densities on the sample surface are reached. A subsequent thermal calculation with the respective RF heat source thus results in a specific sample temperature which is specified in \refT{tab:sim-data-flange} for each excitation magnitude individually. By means of separate calculations the transition from the obtained temperatures to effective powers or even further to parasitic surface resistances is given conclusively for the first three usable quadrupole modes $Q_1$ to $Q_3$.

\begin{figure}[htb]
	\centering
	\setlength{\unitlength}{0.9\columnwidth}
	\includegraphics[width=0.9\columnwidth]{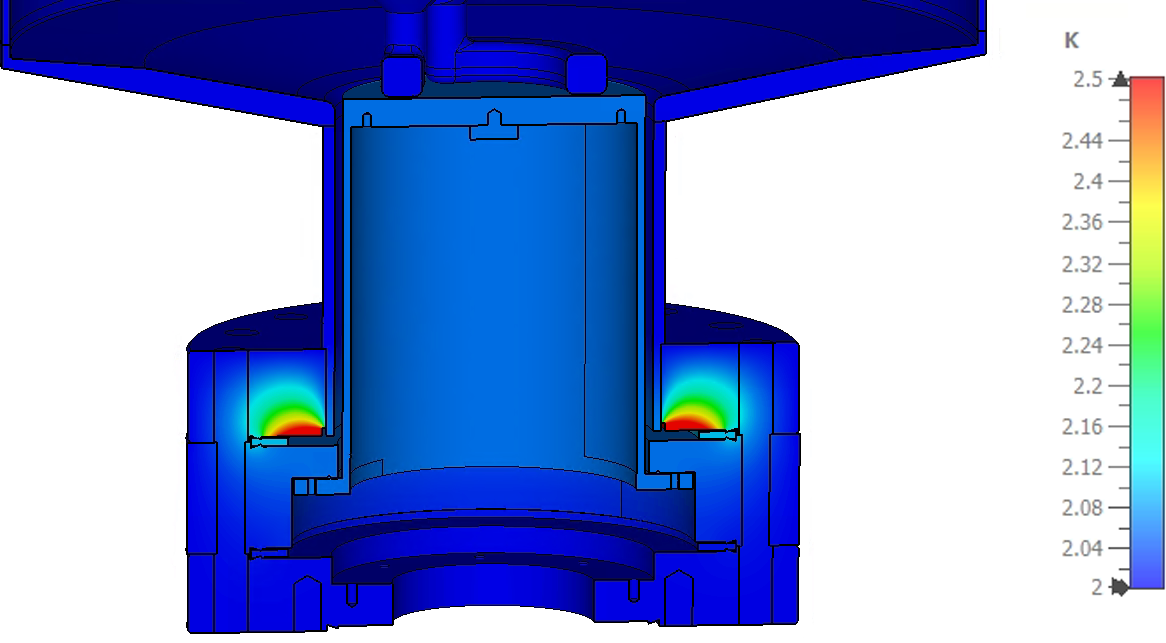}
	\begin{picture}(0,0)(1,0)
		\put(-0.02,0.22){Flange}
		\put(-0.02,0.34){Cavity}
		\put( 0.08,0.20){\line(1,-1){0.08}}
		\put( 0.08,0.38){\line(1, 1){0.08}}
	\end{picture}
	\caption{Simulated temperature distribution due to power dissipation in finite electric conductive materials. The surface of the middle flange corresponding to the mounting of the sample has been coated with niobium.}
	\label{fig:simulation-temperature-mode-sputter}
\end{figure}
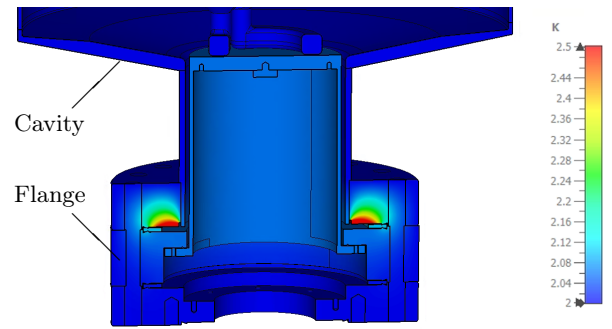

Detailed numerical simulations with normal conductive stainless steel flanges indicate that the unacceptable heating of the sample is mainly attributed to the adapter flange of the sample as can be observed from the temperature distribution visualized in \refF{fig:simulation-temperature-mode}. For this reason, a similar scenario with a modified flange has been investigated. The undesired occurrence of surface heating is prevented by niobium sputtered on the field-exposed surfaces. In the numerical model the electric conductivity is assumed to reach infinity while the thermal conductivity of stainless steel remains unchanged. Compared to the results with the normal conductive flanges the temperature distribution of the modified setup is visualized in \refF{fig:simulation-temperature-mode-sputter}. Here one can clearly see the absence of the strong sample heating while the remaining heating of the upper flange and both seals are still present. The numerical results of the entire simulation cycle are summarized in \refT{tab:sim-data-flange-sputter}.

\section{Sample preparation and flange coating} \label{sec:coating}
The sample tested in this study is a so-called niobium thick film on copper.
A sample was coated with \unit[45]{\micro m} of niobium using DC magnetron sputtering on a bulk copper substrate at INFN Legnaro \cite{vanessa_thick_films_srf_2021}.
The copper surface was prepared by electropolishing and then the niobium film was sputtered in multiple layers (80x \unit[500]{nm} and a top layer of \unit[5]{\micro m}).

The adapter flange used to mount the sample into the QPR is a double-side CF100 flange from 1.4429-ESU (316LN ESR) material.
Prior to the coating, the flange was cleaned in an ultrasonic bath for \unit[10]{minutes}.
The bath was a water-based solution of sodium hydroxide, 1-propanol, and di-sodium tetraborate decahydrate.
After rinsing in deionized water, the flange was dry blown by nitrogen and directly installed into the coating chamber.
On the sample holder the flange was treated by a nitrogen ion-gun and the surface was carefully inspected to be dust free.
To protect the CF flange's knife edge from coating, the area was covered by a metal ring.
This prevents possible particulate creation during the subsequent mounting into the resonator due to potential delamination of the coating from the CF knife edge.

After a bakeout at \unit[650]{\celsius} for \unit[5]{hours}, the base pressure of the coating system dropped to $\unit[1\!\times\!10^{-7}]{hPa}$.
The flange was coated by high power impulse magnetron sputtering (HiPIMS) in Ar atmosphere with a process pressure of $\unit[2.3\!\times\!10^{-2}]{hPa}$, a temperature of \unit[400]{\celsius}, a target power of \unit[600]{W}, and a sample bias of \unit[50]{V\,DC}.
The target was made of a $\unit[90\times100]{mm^2}$ RRR 300 grade niobium plate.
During coating, the flange was continuously moved bidirectionally (rocking) in front of the target to enhance coating thickness uniformity.
The coating process was separated into two parts.
Beginning with a 1.5-hours coating resulting in a \unit[4]{\micro m} thick film, followed by \unit[13]{hours} break under vacuum.
The second coating run started with an annealing at \unit[650]{\celsius} for \unit[5]{hours}, followed by \unit[3]{hours} of coating.
The resultant niobium coating thickness is approximately \unit[12]{\micro m}.
After cooling down, the CF knife edge cover and successively the flange was removed from the machine, inspected optically, packed in standard flange covers and finally sent to be installed into the QPR.

\section{Surface resistance measurements} \label{sec:measurements}

\begin{table}
	\caption{Residual resistance obtained from fitting and selected \trs measurement data at \unit[2.0]{K} and \unit[4.5]{K}.
		For all data given here a constant RF field level of \unit[10]{mT} was applied.}
	\label{tab:data}
	\begin{tabular}{l|c|c|c|c}
		Setup & ~Frequency~ & ~\trs at \unit[2.0]{K}~ & ~\trs at \unit[4.5]{K}~ & \tres \\ \hline \hline
		Baseline & \unit[413]{MHz} & \unit[28.7]{n\ohm} & \unit[110]{n\ohm} & ~\unit[28.5]{n\ohm}~ \\ \hline
		\multicolumn{1}{c|}{niobium} & \unit[417]{MHz}  & \unit[4.7]{n\ohm}  & \unit[74.2]{n\ohm} & \unit[4.2]{n\ohm} \\
		\multicolumn{1}{c|}{coated}  & \unit[844]{MHz}  & \unit[15.2]{n\ohm} & \unit[284]{n\ohm}  & \unit[13.1]{n\ohm} \\
		\multicolumn{1}{c|}{flange}  & \unit[1285]{MHz} & \unit[33.8]{n\ohm} & \unit[696]{n\ohm}  & \unit[31.6]{n\ohm} \\
	\end{tabular}
\end{table}

\begin{figure}[htb]
	\centering
	\includegraphics[width=\columnwidth]{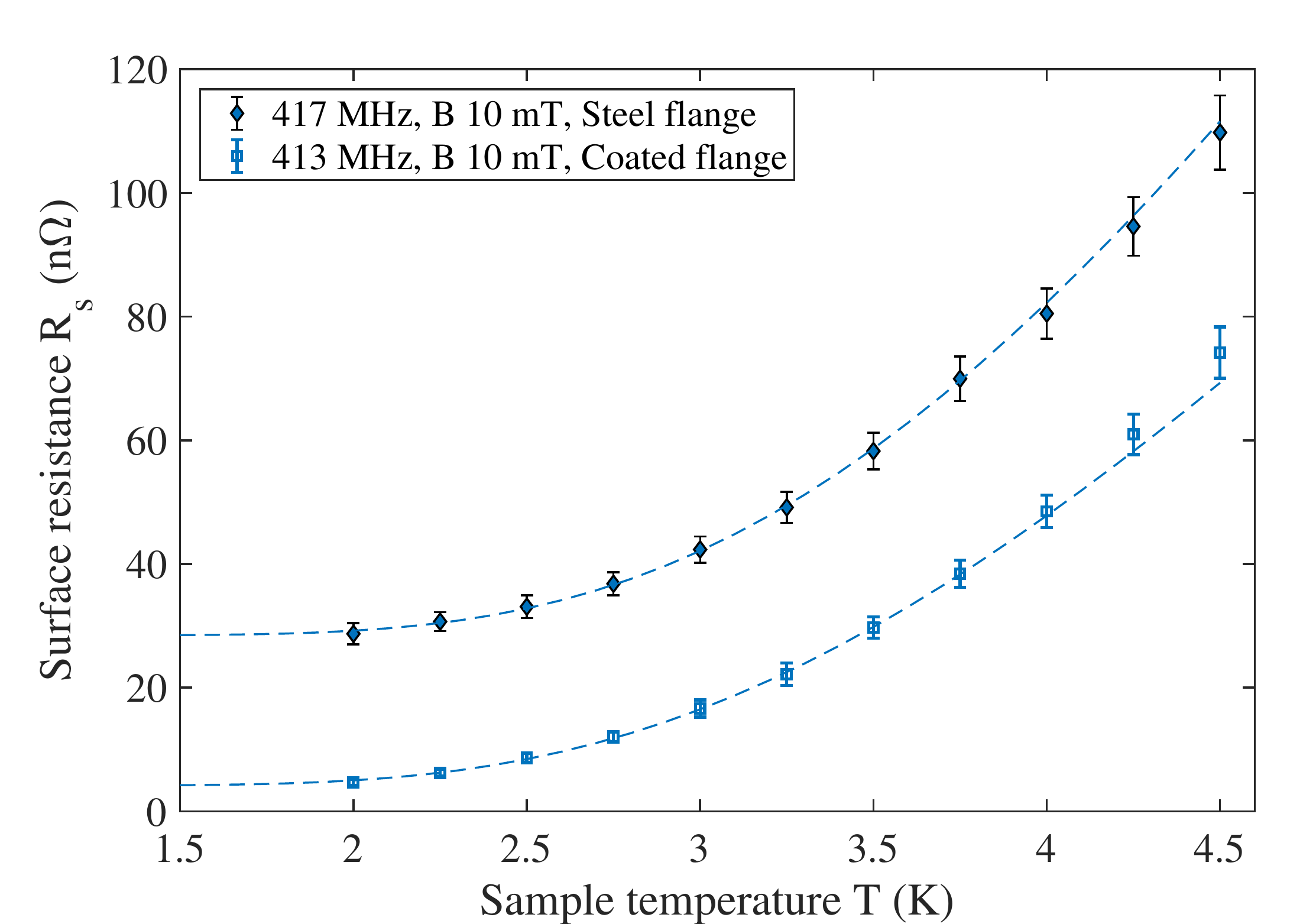}
	\caption{Measured \trs vs.\ sample temperature at the first quadrupole mode for the baseline and the Nb coated flange test. \tres is extrapolated from fits using \refE{eq:rs}. The errorbars represent the combination of systematic error due to RF calibration only with the statistical error due to cryogenic fluctuations and microphonics.}
	\label{fig:rbcs}
\end{figure}

\begin{figure}[htb]
	\centering
	\includegraphics[width=\columnwidth]{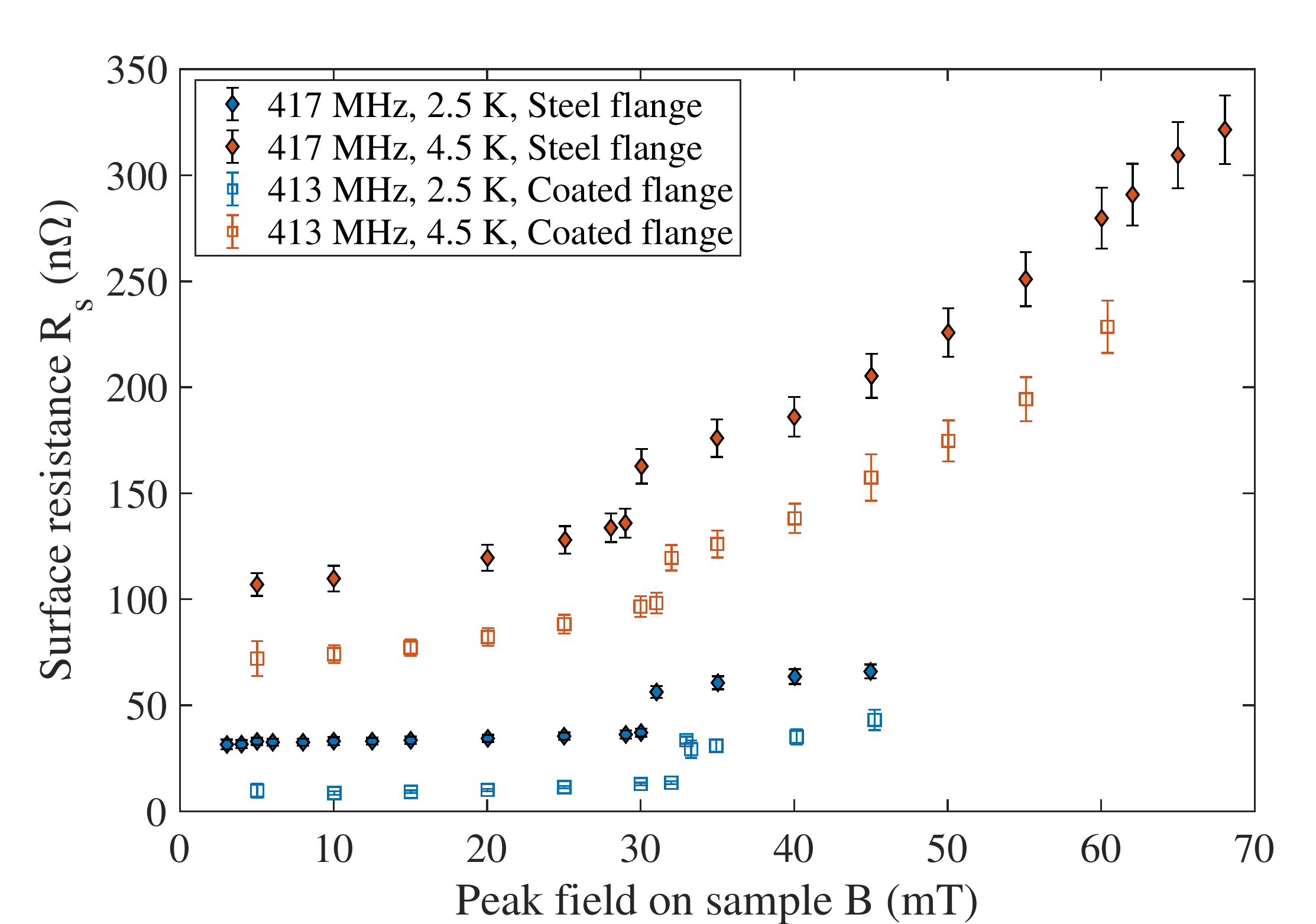}
	\caption{Surface resistance vs.\ RF field results for the baseline and the Nb coated flange test at different temperatures.}
	\label{fig:RsvsB}
\end{figure}

Prior to coating the adapter flange, a baseline test of the sample is conducted.
Measurement data are shown in Figs. \ref{fig:rbcs} and \ref{fig:RsvsB}, together with data of the coated flange.
For data taken at a constant level of RF field (see \refF{fig:rbcs}), \tres is extrapolated using \refE{eq:rs}.
Fit results for baseline measurement and the test with niobium coated flange are given in \refT{tab:data}.
Note that thanks to the low frequency of about \unit[415]{MHz} and sample temperatures down to \unit[2]{K} the uncertainty in \tres is less than \unit[1]{n\ohm}.
Measured values for \trs of about \unit[25-30]{n\ohm} at the first quadrupole mode near \unit[415]{MHz} are typical values, also compared to other samples (see \refF{fig:example-data}).

Measuring surface resistance versus RF field at constant temperature shows a ``jump'' in \trs at about \unit[30]{mT} (see \refF{fig:RsvsB}).
The field level at which the jump occurs depends weakly on temperature, while the amplitude remains constant.
This can be interpreted as a ``Q-switch'' behavior of the sample, independent of the adapter flange \cite{padamsee_srf_accelerators}.
The visible ``jump'' in \trs data at the same field level and with the same amplitude for baseline and Nb coated flange test excludes significant errors coming from the RF measurements or possible mounting issues.

\begin{figure}[htb]
	\centering
	\includegraphics[width=\columnwidth]{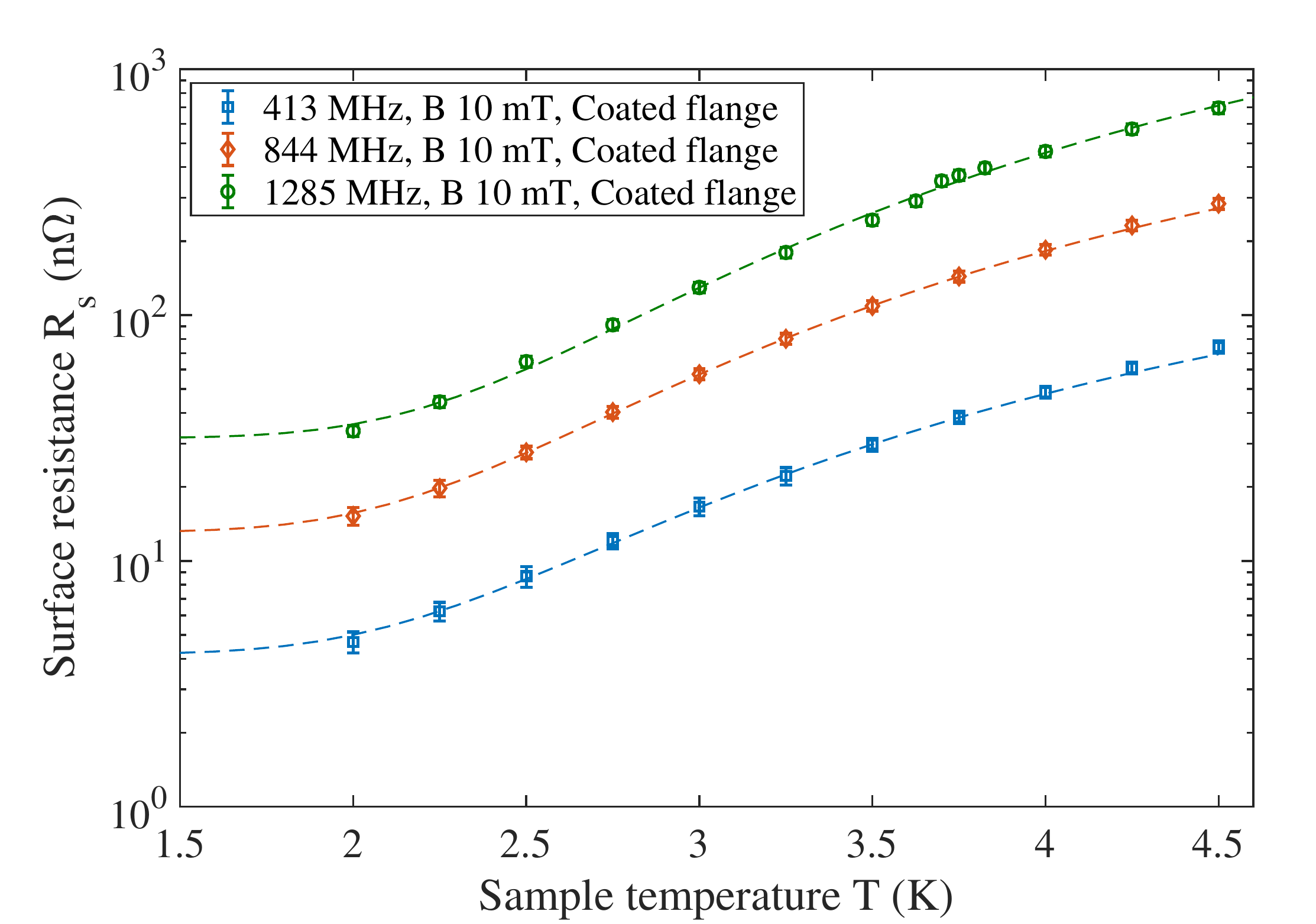}
	\caption{Measured \trs vs.\ sample temperature for all three quadrupole modes, all measured with the niobium coated flange at \unit[10]{mT}. \tres is extrapolated from fits using \refE{eq:rs}.}
	\label{fig:rvst-allmodes}
\end{figure}

Thanks to the strong suppression of RF dissipation on the niobium coated adapter flange, especially at higher frequencies, measurements of \trs were possible at all three quadrupole modes at temperatures down to \unit[2]{K}.
Measurement data together with fits extrapolating \tres according to \refE{eq:rs} is shown in \refF{fig:rvst-allmodes}.

\section{Discussion} \label{sec:discussion}
The presented measurements clearly show that the application of a superconducting niobium coating on the stainless steel adapter flange at the far end of the coaxial gap between quadrupole resonator and QPR sample solved the problem with systematic errors from which the QPR suffered in the past.
The surface resistance versus temperature measurement of the first quadrupole mode with the Nb-coated adapter flange (see \refF{fig:rbcs}) shows a reduction by \unit[24]{n\ohm} over the entire temperature range.
The numerical simulations predicted only \unit[12.6]{n\ohm} for this mode, see table \ref{tab:sim-data-flange};
however, the simulation was performed with an idealized geometry and without allowing for geometrical tolerances.
The difference might stem from such deviations, for example from coaxiality or co-planarity between pole-shoes and sample.
They result in a larger fraction of dipole mode or even give rise to a monopole mode to propagate through the coaxial gap reaching the bottom flange.
The obtained results suggest, that such conceivable problems were also solved by coating the adapter flange with niobium.

The measured \trs values are now comparable with the ones obtained in SRF cavity measurements.
Up to now, no sample-test cavity has demonstrated absolute \trs values lower than \unit[5]{n\ohm}.
The measurements of \trs as a function of the RF magnetic field show that the error in \trs\ -- originating from parasitic losses in the stainless steel adapter flange -- is nearly independent of temperature and applied RF field strength (see \refF{fig:RsvsB}).
This was expected from the numerical simulations because the dissipated power depends quadratically on the RF field amplitude for both, superconductors and normal conductors.
Hence, the contribution of parasitic losses to the observed \trs mainly affects the measurement accuracy of \tres and can be treated as a systematic bias. On a side note, the measurement precision, i.e.\ the reproducibility of a measurement has always been very good, and was not further improved by the flange coating.
It is conceivable that a dependence on temperature or RF field amplitude could originate from temperature dependent thermal conductivities, however, this would only be a second-order effect and could not explain the measured data.

Looking at the simulation results for the third quadrupole mode (Q3) in the case of normal conducting adapter flange sheds light on the question of a temperature dependent bias from parasitic losses (see \refT{tab:sim-data-flange}).
Increasing the RF field amplitude up to \unit[100]{mT}, which is close to the quench limit of the QPR, leads to an increased bias of \unit[74]{n\ohm}, compared to \unit[63]{n\ohm} at low field.
This difference of about \unit[11]{n\ohm} corresponds to a sample temperature of \unit[6.5]{K} as opposed to \unit[2]{K} at low field.
Hence, when extracting BCS parameters for niobium at temperatures up to \unit[4.5]{K}, the temperature dependent component of the bias will be smaller than \unit[11]{n\ohm}.
Due to the strong increase of \tbcs with temperature and a minimum RF measurement uncertainty of about \unit[5]{\%}, BCS fit results will not be affected significantly by a possibly temperature dependent bias.

Unfortunately, baseline data for this sample is not available at higher harmonic quadrupole modes.
However, operational experience indicates that the reduction of the measured \trs is significantly higher at higher frequencies.
Especially at the third mode, \tres has never been lower than several hundred nano-ohm.
Here, the absolute numbers of expected \trs bias from parasitic heating as obtained from the numerical simulations seem low.
Analogous to the considerations of the first quadrupole mode this might be due to dipole mode components of the RF field penetrating the coaxial structure and heating a normal conducting flange.
Simulations investigating this effect are ongoing, first results show that indeed manufacturing tolerances and geometrical imperfections lead to an increased level of RF field at the end of the coaxial structure.
This would lead to additional parasitic heating that is not considered in the simulations discussed here, and is also prevented by the niobium coating of the flange.
Especially dipole components at the third quadrupole mode excited above the dipole cutoff frequency would cause a significant contribution when dissipating on normal conducting surfaces.

\section{Summary} \label{sec:summary}
In this paper we present an extensive numerical study quantifying the impact of parasitic losses at a normal conducting adapter flange for QPR measurements.
Coating those areas of the stainless steel flange that are exposed to RF fields with several micrometer of niobium strongly suppresses the systematic bias in surface resistance measurement.
This bias mainly affected the extracted residual resistance.

Measurements with a niobium thick-film sample before and after coating the adapter flange showed a reduction of measured surface resistance by \unit[24]{n\ohm} at the first quadrupole mode.
While it depends only weakly on temperature and the RF field amplitude there is a strong frequency dependence.
At the third quadrupole mode surface resistance values lower than \unit[35]{n\ohm} have been achieved, boosting the accuracy of the measurement system to an unprecedented level.
Establishing coated flanges as a new standard for QPR samples opens up new possibilities for the RF characterization of superconducting samples, \eg investigating frequency dependent residual resistance.

\begin{acknowledgments}
We thank V.~Garcia Diaz, E.~Chyhyrynets and C.~Pira (INFN Legnaro, Italy) for supplying the niobium thick-film QPR sample which has received funding from the European Union's Horizon 2020 Research and Innovation programme under Grant Agreement No 101004730. % i.FAST

We would like to express our gratitude to the SMART collaboration -- coordinated by the spokesperson W.~Hillert (Universit\"{a}t Hamburg) -- for many fruitful discussions.
This research has been funded by the Federal Ministry of Education and Research of Germany (BMBF), project number 05K2019 - SMART.
\end{acknowledgments}

\section*{Data Availability}
The data that support the findings of this study are available from the corresponding author upon reasonable request.

% Create the reference section using BibTeX:
%\bibliography{literature}

%aipnum4-2.bst 2019-01-14 (MD) hand-edited version of apsrev4-1.bst
%Control: key (0)
%Control: author (8) initials jnrlst
%Control: editor formatted (1) identically to author
%Control: production of article title (0) allowed
%Control: page (1) range
%Control: year (1) truncated
%Control: production of eprint (0) enabled
%

\end{document}